\documentclass[twocolumn,twoside,slac_two]{revtex4}
\usepackage{graphicx}
\usepackage{fancyhdr}
\begin{document}

%
\def \ie    {\hbox{\it i.e.}}     
\def \etc   {\hbox{\it etc.}}
\def \ibid  {\hbox{\it ibid.}}
\def \vs    {\hbox{\it vs.}}
\def \eg    {\hbox{\it e.g.}}     
\def \cf    {\hbox{\it cf.}}
\def \etal  {\hbox{\it et al.}}
\def \via   {\hbox{\it via}}
\def \CPC   {Comput. Phys. Commun.~}
\def \EPJ   {European Phys. Journal~}
\def \JHEP  {J. High Energy Phys.~}
\def \NIM   {Nucl. Instr. Meth.~}
\def \NP    {Nucl. Phys.~}
\def \PL    {Phys. Lett.~}
\def \PRD   {Phys. Rev. D~}
\def \PRL   {Phys. Rev. Lett.~}
\def \PRep  {Phys. Reports~}
\def \RMP   {Rev. Mod. Phys.~}
\def \zphys {Z. Phys.~}
\hyphenation{back-ground}
\hyphenation{brem-sstrah-lung}
\hyphenation{cal-or-ime-ter cal-or-ime-try}
\hyphenation{had-ron had-ronic}
\hyphenation{like-li-hood}
\hyphenation{posi-tron posi-trons}
\hyphenation{semi-lep-tonic}
\hyphenation{syn-chro-tron}
\hyphenation{system-atic}
%
%
\def \beq   {\begin{equation}}
\def \eeq   {\end{equation}}
\def \bbeq  {\begin{eqnarray*}}
\def \ebeq  {\end{eqnarray*}}
\def \bbeqn  {\begin{eqnarray}}
\def \ebeqn  {\end{eqnarray}}
\def \Tr    {\mathop{\mathrm Tr}}
\def \Im    {\mathop{\mathrm Im}}
\def \Re    {\mathop{\mathrm Re}}
\def \vect  {\overrightarrow}
\def \twdl  {\widetilde}
\def \hat   {\widehat}
\def \partder#1#2  {\partial #1\over\partial #2}
\def \secder#1#2#3 {\partial^2 #1\over\partial #2 \partial #3}
%
%
\def \omg#1 {\mbox {${\mathcal O}(#1)$}}
\def \avg#1 {$\left\langle #1\right\rangle$}
\def \to    {\rightarrow}
\def \bra#1 {$\left\langle #1\right|$}
\def \ket#1 {$\left| #1\right\rangle$}
\def \braket#1#2 {\left\langle #1\right. \left| #2\right\rangle}
\def \amp#1 {${\mathcalA}(#1)$}
\def \apgt  {\raisebox{-0.6ex}{$\stackrel{>}{\sim}$}}
\def \aplt  {\raisebox{-0.6ex}{$\stackrel{<}{\sim}$}}
\def \pma#1#2 {\mbox{\raisebox{-0.6ex}
           {$\stackrel{\scriptstyle \;+\; #1}{\scriptstyle \;-\; #2}$}}}
%
\def \dr    {$\;^\mid\!\!\!\longrightarrow$}
%
\def \ev    {\,\mathrm {eV}}
\def \kev   {\,\mathrm {keV}}
\def \mev   {\,\mathrm {MeV}}
\def \gev   {\,\mathrm {GeV}}
\def \tev   {\,\mathrm {TeV}}
\def \km    {\,\mathrm {km}}
\def \cm    {\,\mathrm {cm}}
\def \mm    {\,\mathrm {mm}}
\def \um    {\,\mu\mathrm m}
\def \ghz   {\,\mathrm {GHz}}
\def \mhz   {\,\mathrm {MHz}}
\def \khz   {\,\mathrm {kHz}}
\def \ps    {\,\mathrm {ps}}
\def \ns    {\,\mathrm {ns}}
\def \us    {\,\mu\mathrm s}
\def \ms    {\,\mathrm {ms}}
\def \hz    {\,\mathrm {Hz}}
\def \pb    {\,\mathrm {pb}^{-1}}
\def \fb    {\,\mathrm {fb}^{-1}}
\def \mrad  {\,\mathrm {mrad}}
\def \BR#1#2 {\mbox{Br}(#1$\to$#2)}
\def \JP     {\mathrm J$^{\mathrm P}$}
\def \Mw    {${\mathrm M}_W$}
\def \Mz    {${\mathrm M}_Z$}
\def \Mpi   {${\mathrm M}_\pi$}
\def \Mk    {${\mathrm M}_K$}
\def \Gf    {G$_{\mathrm F}$}
\def \As    {$\alpha_s$}
\def \Mt    {${\mathrm M}_{t}$}
\def \Mb    {${\mathrm M}_{b}$}
\def \Mc    {${\mathrm M}_{c}$}
\def \Ms    {${\mathrm M}_{s}$}
\def \Mud   {${\mathrm M}_{u,d}$}
\def \Mh    {${\mathrm M}_{H}$}
\def \sW    {$\sin^2\theta_{\mathrm W}$}
\def \sWeff {$\sin^2\theta_{\mathrm W}^{eff}$}
\def \gV    {g$_V$}
\def \gA    {g$_A$}
\def \lms   {\Lambda_{\overline{\mathrm MS}}}
\def \Vub   {$|$V$_{\mathrm{ub}}|$}
\def \Vus   {$|$V$_{\mathrm{us}}|$}
\def \Vcb   {$|$V$_{\mathrm{cb}}|$}
\def \MET   {\mbox{$E_T\hspace{-1.2em}\slash\hspace{1.0em}$}}
\def \PET   {\mbox{$p_T$}}
\def \pT    {\mbox{$p_\perp}}
\def \dedx {d{\it E}/d{\it x}}
\def \rphi {$r$-$\phi$}
\def \Pt2  {${\mathrm P}_\perp^2$}
%
\def \epem {$e^+e^-$}
\def \mpmm {$\mu^+\mu^-$}
\def \tptm {$\tau^+\tau^-$}
\def \ppb  {$p\overline p$}
\def \ttb  {$t\overline t$}
\def \lplm {$\ell^+ \ell^- $}
\def \J    {$\mathrm J/\psi$}
\def \Ks   {$\mathrm K^0_{\mathrm S}$}
\def \Kl   {$\mathrm K^0_{\mathrm L}$}
\def \Bs   {$\mathrm B_{\mathrm S}$}
\def \Bo   {$\mathrm B^0$}
\def \Bp   {$\mathrm B^+$}
\def \Ds   {$\mathrm D_{\mathrm S}$}
\def \Do   {$\mathrm D^0$}
\def \Dp   {$\mathrm D^+$}
\def \Bss  {$\mathrm B_{\mathrm S}^*$}
\def \Bso  {$\mathrm B^{*0}$}
\def \Bsp  {$\mathrm B^{*+}$}
\def \Dss  {$\mathrm D_{\mathrm S}^*$}
\def \Dso  {$\mathrm D^{*0}$}
\def \Dsp  {$\mathrm D^{*+}$}


\title{Beyond Standard Model Physics}

%

\author{Leo Bellantoni}
\affiliation{Fermi National Accelerator Laboratory, Batavia, IL 60510, USA}

\begin{abstract}
There are many recent results from searches for fundamental new physics using
the TeVatron, the SLAC $b$-factory and HERA.  This talk quickly reviewed
searches for pair-produced stop, for gauge-mediated SUSY breaking, for Higgs
bosons in the MSSM and NMSSM models, for leptoquarks, and v-hadrons.  There
is a SUSY model which accommodates the recent astrophysical experimental results 
that suggest that dark matter annihilation is occurring in the center of our
galaxy, and a relevant experimental result.  Finally, model-independent searches
at D0, CDF, and H1 are discussed.
\end{abstract}

\maketitle

\thispagestyle{fancy}

\section{Introduction
		\label{sec:Intro}}

It is somewhat misleading to use the terms 'New Physics' or 'Beyond Standard
Model Physics' for results that explicitly search for signatures that would
result from extensions to the standard model.  In almost any precision
measurement of the production, properties or decays of already-known property,
there is the possibility of an unusual result can be explained with an 
extension to the standard model.  In a certain sense then, nearly every result
being shown here this week is a search for new physics.  In particular, 
Lee Roberts will be giving a presentation on searches for 'New Physics' in
low energy experiments.

However even restricting the discussion to analyses that search directly for
extensions to the standard model at high energy colliders would create a
discussion that goes on for far too long.  With great regret, I have had to
trim my selection of topics very sharply.  There is just too much good work
done here to give each study adequate coverage.  In particular, there has been
a great deal of excellent work preparing for the analysis of the imminent
flood of data from the LHC.  I can only refer the reader to the parallel
sessions of this conference.

The bulk of the recent results are from the TeVatron, which has been performing
quite well.  Although most of the results described here are based on smaller
datasets, just under $7\fb$ have been delivered to each experiment 
at this time.  Results from the CDF and D0 collaborations are collected
at~\cite{TeV_web}.

Certain characteristics of hadron collisions are common to all or nearly all
searches for exotic phenomena in them.  The copious production of multijet
``QCD" events is suppressed with detectors designed to reject fake electrons
and muons (hereafter referred to as leptons, $\ell$), kinematic cuts and
isolation cuts that require that the identified lepton not be surrounded by
other activity in the detector.  Multijet background is rarely, if ever,
modeled effectively with Monte Carlo simulation techniques.  For the many
searches which select highly energetic leptons or momentum imbalance in the
final state, the following known physics processes typically produce significant 
backgrounds: the Drell-Yan process \ppb$\to \gamma^* / Z \to$\lplm,
$\gamma^* / Z \to$\tptm, $W^\pm \to \ell^\pm\nu$, and \ttb~production.  The
diboson production processes \ppb$\to VV$ with $V \in \{\gamma, Z,W\}$  have 
lower production cross-sections but also create unusual signatures which are of
interest in many searches.

sJust as limiting as backgrounds are the kinematic facts of life in hadron
colliders.  In \ppb~(or $pp$) collisions, the component of the initial-state
momentum along the collision axis is not known and kinematic calculations can
only be done in the plane perpendicular to the collision.  I will use \MET
to indicate the opposite of the observed sum of particle momenta in this
transverse plane, and \PET~to indicate the momentum of an object projected onto
this transverse plane.

\section{About Supersymmetry
		\label{sec:SUSY}}

As many of the results discussed here are based on supersymmetric (SUSY)
extensions of the standard model, a short introduction is appropriate.  In no
way however can this replace the many excellent existing reviews and
introductions, some of which may be found in 
reference~\cite{SUSY_reviews,Martins}.

SUSY provides solutions to several existing dilemmas in the standard model.
One is the ``\Mh problem".  The propagator for a Higgs scalar with fermionic
couplings $\mathcal{L} = - \lambda_f H f {\bar f}$ has one loop correction terms 
that contribute to the mass in amount
$\Delta {\mathrm M}_H^2 = -\frac{|\lambda_f|^2}{8\pi^2} \Lambda_{UV}^2$, where
$\Lambda_{UV}$ is a cutoff scale corresponding to the point where our
existing understanding of nature's particle content becomes inadequate.
Our difficulty is that we have no clear value for $\Lambda_{UV}$ short of the
Plank scale, resulting in large negative contributions to $m_H$.  However,
if for every fermion there is a corresponding scalar field $S$ with interaction
$\mathcal{L} = -\lambda_s |h|^2 |S|^2$, then the corresponding scalar loop
diagram introduces canceling mass contributions
$\Delta _H^2 = \frac{\lambda_S}{16\pi^2} [ \Lambda_{UV}^2 + \ldots ]$.

A second outstanding problem in the standard model is the dark matter problem.
The lightest neutral sparticle often makes a good dark matter candidate.

Finally, the coupling constants for the strong, weak and electromagnetic forces 
vary with the energy scale of the interaction according to the renormalization
group.  In the minimal supersymmetric extension to the standard
model (MSSM), the coupling constants evolve out to reach similar values at the
scale of $10^{16}\gev$, which does not happen in the standard model.
Quoting~\cite{Martins}, ``While the apparent unification of gauge couplings at
[this scale] might just be an accident, it may also be taken as a strong hint
in favor of a grand unified theory or superstring models, both of which can
naturally accommodate gauge coupling unification below ${\mathrm M}_P$."

Because the SUSY mass spectrum evidently differs from that of the standard
model particle content, there must be SUSY-breaking terms in the Lagrangian.
The primary constraint on these terms is that they not reintroduce ultraviolet
divergences of the sort we were glad to be rid of earlier.  This is not a
very tight constraint; there are at least 105 new free parameters in the most
general form of the symmetry breaking Lagrangian.  What this does is provide
a flexible framework in which different models of symmetry breaking can be
inserted and investigated.  Of the many different physical concepts that can
and have been inserted into the SUSY breaking Lagrangian, two of the most 
studied ones are the mSUGRA and the GMSB models.  We have recent results in
both of these SUSY breaking models.

The generality of the SUSY breaking Lagrangian is perhaps why the SUSY
hypothesis has had such a long run.  After all, SUSY was proposed in the early
1970s, when the standard model was still a novel model, and much of its particle
content was unknown.  For nearly 4 decades, theorists have been able to write
Lagrangians of all sorts into this framework and work out the their possible
implications.

$R$-parity is a hypothesized quantum number which differentiates standard
model particles from SUSY particles.  All of the searches presented here
assume the conservation of $R$-parity, so that each SUSY particle is produced
in conjunction with the corresponding SUSY anti-particle.

In SUSY, 2 Higgs doublets
\beq
H_d = \left( \begin{array}{c} H_d^0 \\
							  H_d^- \end{array} \right)
\hspace{15mm}
H_u = \left( \begin{array}{c} H_u^+ \\
							  H_u^0 \end{array} \right)
\label{Eqn:Higgses}
\eeq
coupling respectively to down- and up- type fermions are required in order to
prevent triangle anomalies.  The ratio of the vacuum expectation values of
the two neutral fields,
$\tan \beta = \langle H_u^0 \rangle / \langle H_d^0 \rangle$
is one of the key parameters of supersymmetry, or indeed of any 2 doublet
model.  Analyses that apply Bayesian methods to a random samplings of parameter
space~\cite{Bayes} strongly favor larger values of $\tan \beta$ at least in the
context of mSUGRA and similar SUSY-breaking models.

The SUSY partners to the Higgs fields are the spin 1/2 Higgsinos:
\beq
\tilde{H}_d = \left( \begin{array}{c} \tilde{H}_d^0 \\
									  \tilde{H}_d^- \end{array} \right)
\hspace{20mm}
\tilde{H}_u = \left( \begin{array}{c} \tilde{H}_u^+ \\
									  \tilde{H}_u^0 \end{array} \right)
\label{Eqn:Higgsinos}
\eeq
The charged components of the Higgsino fields can form linear admixtures
with the wino to create 2 charginos, $\tilde{\chi}^\pm$.  The
convention is that $m(\tilde{\chi}^\pm_1) < m(\tilde{\chi}^\pm_2)$.  
$\tilde{\chi}$, without subscript, refers the lightest of the mixtures.
The neutral components of the Higgsino fields form linear admixtures
with the zino and photino to create 4 neutralinos, $\tilde{\chi}^0_i$.

Returning to the scalars, after electroweak symmetry breaking, two doublet
models yield 5 Higgs bosons: two $CP$-even neutral scalars $h$ and $H$, a
$CP$-odd neutral $A$ and a pair of charged scalars, $H^\pm$.

No discussion at length about SUSY is complete without mentioning that the
MSSM at least is under some pressure from experimental results from the
electroweak symmetry breaking sector.  The lightest neutral MSSM Higgs boson
$h$ must have a mass below 135 GeV~\cite{Martins,Pesky} and experimental lower
bounds~\cite{LEP_Higgs} have come to approach this level.

\section{Searches for $\tilde{t}$
		\label{sec:stop}}

We have recent search results for the pair production of the SUSY partner to
the top quark in \ppb~ collisions.  There are 3 decay channels under study.
In all 3 cases, limits are placed in a plane where the horizontal axis is the
mass of the pair-produced $\tilde{t}$ and vertical axis is the mass of the
final state SUSY particle.

The first channel is
$\tilde{t} \to b \tilde{\chi}^+; \tilde{\chi}^+ \to \tilde{\nu} \ell^+$.
$R$-parity conservation means that the charge conjugate process occurs on the
other side of the event, where a $\overline{\tilde{t}}$ decays similarly.  The
signature is an \lplm~pair with \MET from the escaping neutrinos.  There are 2
$b$-jets in the final state, but both the D0 and CDF collaborations found
kinematic selection sufficient.  The recent D0 result~\cite{D0_stop_emu} uses
$3.1\fb$ in the $e-\mu$ channel in conjunction with earlier
$1.1\fb$ $e-\mu$ and $e-e$ results.  The CDF result~\cite{CDF_stop_dilep} is
based on $1.0\fb$ in all 3 dilepton channels.  Limits are drawn in the
$m(\tilde{\nu})$ \vs~$m(\tilde{t})$ plane and extend up to
$m(\tilde{\nu}) \simeq 120\gev$.

The second channel is
$\tilde{t} \to b \tilde{\chi}^+; \tilde{\chi}^+ \to \tilde{\chi}^0 (W^+/H^+/G^+)$
where the remaining charged gauge boson decays semileptonically.  This channel
was originally of interest when measured values of $m(t)$ seemed to be a little
lower in the dilepton channel.  It is possible if $m(\tilde{t}) < m(t)$ for the
SUSY process to contaminate the \ttb~dilepton channel and pull down the 
apparent $t$  quark mass.  With 4 undetected particles in the final state
(the $\tilde{\chi}^0$ is taken to be stable), the kinematics are very
underconstrained, even in the transverse plane.  However, one may use a
weighted sum of possible solutions to the kinematic problem to estimate
$m(\tilde{t})$.  CDF has set limits~\cite{CDF_Robin} on $m(\tilde{\chi}^0)$
as a function of $m(\tilde{t})$ (up to $197\gev$) and the assumed
\BR{$\tilde{\chi}^\pm$}{$\tilde{\chi}^0\nu\ell^\pm$} ~using $2.7\fb$.

The third channel to be studied is $\tilde{t} \to c \tilde{\chi}^0$.  As the
lifetime of charm hadrons typically is shorter than that of bottom hadrons,
and as the transverse momentum of the charged products of charm decays
typically is less than that of bottom decays, obtaining a pure sample of charm
decays with impact parameter tagging is very difficult.  The CDF collaboration
has developed a 2 output, 22 input neural network that distinguishes (at one
output) between charm and bottom jets.  The other output distinguishes between
charm and light or $\tau$ jets.  Cutting on the sum of the two outputs, they
set limits~\cite{CDF_underdocumented} in the $m(\tilde{\chi}^0)$
\vs~$m(\tilde{t})$ plane extending up to $m(\tilde{t}) = 180\gev$.

\section{Trifermion SUSY Searches
		\label{sec:threebees}}

\begin{figure}[h]
\centering
\includegraphics[width=80mm]{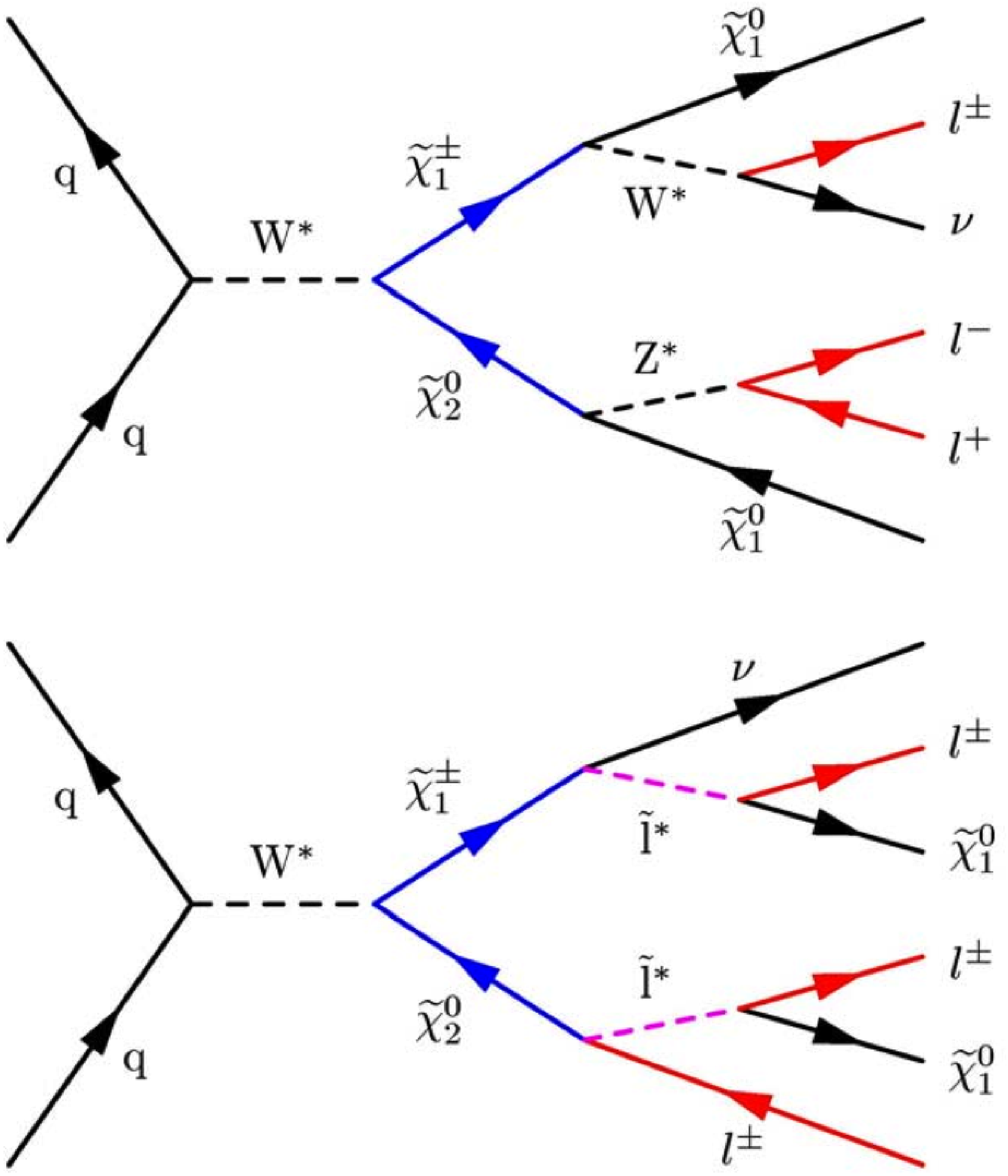}
\caption{Some of the ways in which SUSY creates trilepton signatures in
\ppb~collisions.}
\label{Fig:trileptons}
\end{figure}

SUSY allows a number of channels leading to 3 leptons in the final state,
as shown in Figure~\ref{Fig:trileptons}.  There are relatively few backgrounds,
but the cross-section for production times the branching ratio for decay into
any particular combination of leptons is small.  Depending on the particular
values of the SUSY-breaking parameters (here, the mSUGRA breaking is used) it
may happen that the mass of the charginos or neutralinos produced at the
$q \overline q$ vertex is only a little larger than the mass of the escaping 
$\tilde{\chi}_1^0$, in which case a low momentum lepton is produced.  For the
high values of $\tan \beta$ that are of particular interest, $\tau^\pm$ leptons
are often produced, and are detected by their decays to electrons or muons that
are also of lower momentum.  To increase sensitivity then, it is common to not
attempt to identify the lepton of third lowest \PET, but rather to just ask for
a charged particle that is isolated from any jets that appear in the event.
Robert Forrest and Todd Adams have presented the CDF~\cite{CDF_trileptons} and
D0~\cite{D0_trileptons} results in this conference.

Another final state with three fermions produced via SUSY diagrams has been
investigated by CDF~\cite{CDF_lljj}.  Suppose that in the top diagram of
Figure~\ref{Fig:trileptons} the $W$ materializes as a $q\overline q$ pair,
creating 2 jets.  The resulting event then appears as a $WZ$ pair with \MET.
In the standard model, hadronically decaying $W$s with $Z\to$\lplm do not have
\MET except as a result of mismeasurement, so this is a relatively clean final
state.  While it has to be admitted that the existing sensitivity is not really
comparable to what might reasonably be expected in SUSY, there are several
reasons why large improvements can be expected in the future. To date, only
$Z\to$\epem has been investigated, and $b$-jet identification has not been 
employed although a large \ttb~background is present.  Also, the present result
is based on $2.7\fb$ of data at one of the two TeVatron experiments; a final
sample some 7 or 8 times larger than this could occur.

\section{Gauge Mediated Supersymmetry Breaking
		\label{sec:GMSB}}

In order to give different mass spectra to SUSY \vs~standard model particles
using gauge interactions, one can postulate the existence of new fields, called
messengers, that couple the standard model and SUSY particles to an ultimate
source of symmetry breaking.  In these GMSB models, the lightest neutral SUSY
particle is nearly always the gravitino, which is an interesting dark matter
candidate for masses on the scale of a few $\kev$.  For the collider
experimentalist, the way to think of various versions of this model is to
categorize them in terms of their next-to-lightest SUSY particle (NLSP).
Whatever particular SUSY particles might be created at the hard scattering
vertex, they will cascade down to the NLSP (assuming $R$-parity conservation)
which will after some lifetime go to an undetected gravitino.  The nature of
the NLSP will then determine what type of events to look for in the dataset. 

\begin{figure}[h]
\centering
\includegraphics[width=80mm]{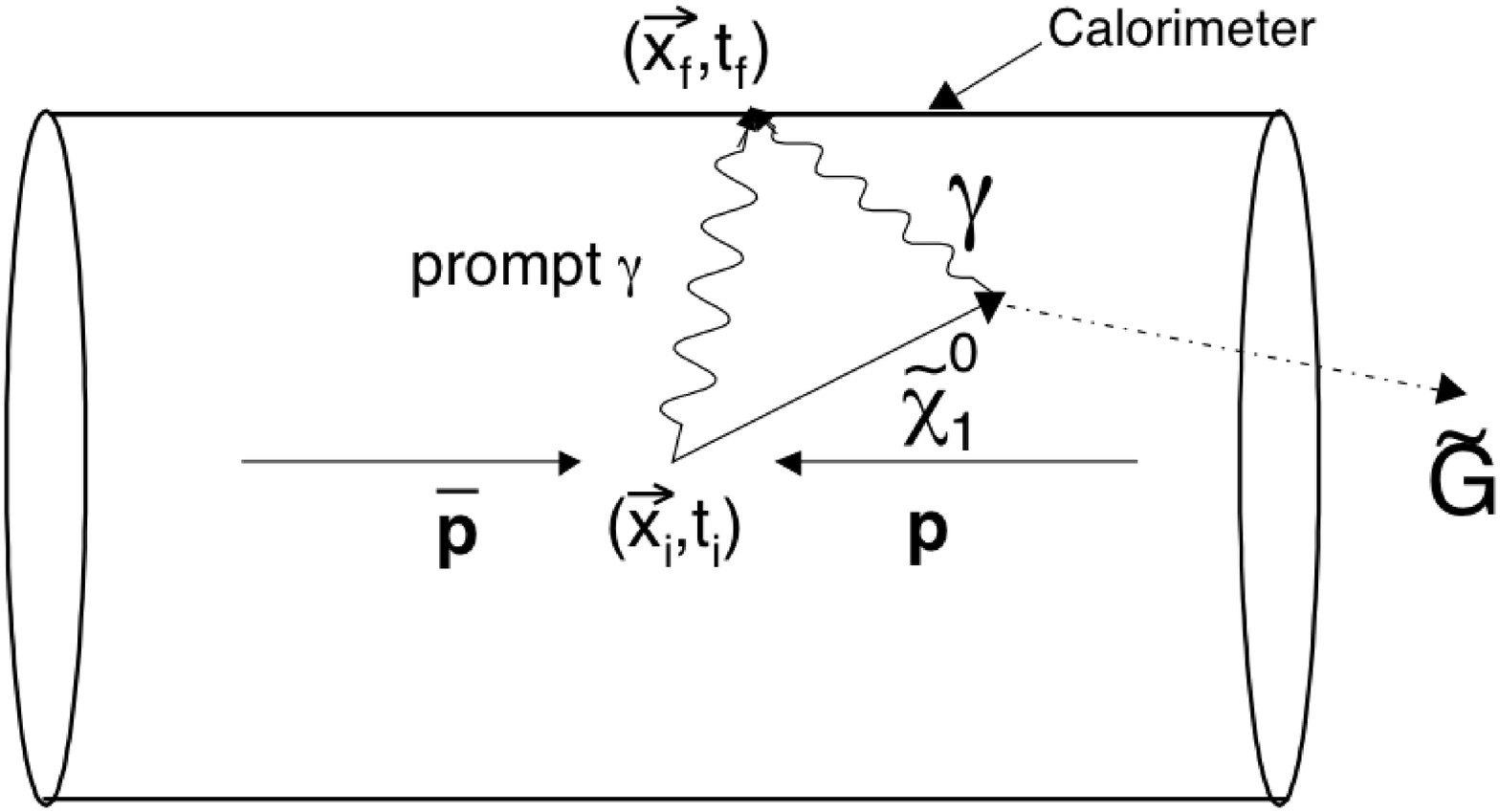}
\caption{Why photons from $\tilde{\chi}^0 \to \gamma\tilde{G}$ arrive late
in the electromagnetic calorimeter of a large collider experiment.}
\label{Fig:latelight}
\end{figure}

When the NLSP is the lightest neutralino and $m(\tilde{\chi}^0) <$\Mz, its
decay produces a photon in conjunction with the gravitino.  If the 
$\tilde{\chi}_1^0$ lifetime is on the order of $10\ns$, the arrival of the
$\gamma$ will be delayed because of the flight path, as shown in
Figure~\ref{Fig:latelight}.  The CDF detector has $~0.5\ns$ time resolution in
its EM calorimeter, which makes this type of search feasible.  In addition to
the delayed photon, the search requires a jet and \MET.  Limits up to
$m(\tilde {\chi}^0) > 191\gev$ for $\tau(\tilde {\chi}^0) > 5\ns$ were obtained
and a detailed description of the analysis was published in
2008~\cite{CDF_latelight}.

When the lifetime of the neutralino is on the order of a few $\ns$ or less,
the delayed photon technique will not work.  However, as a consequence of
$R$-parity, there should be 2 SUSY cascades in the event leading to 2
NLSP $\tilde{\chi}^0$ decays to $\gamma\tilde{G}$.  In this conference,
Eunsin Lee has reported on a search~\cite{CDF_earlylight} for GMSB at CDF which
requires 2 photons with high \PET~along with \MET from the gravitinos.  Limits
up to  $m(\tilde {\chi}^0) > 149\gev$ for $\tau(\tilde {\chi}^0) < 1\ns$ were
obtained.

\section{MSSM Higgs
		\label{sec:MSSM_Higgs}}

In the large $\tan \beta$ limit, the mass and couplings of the $A$ boson
approach the mass and couplings of one of the two $CP$-even bosons $h$ or $H$.
If $A \to H$ and $m(H)$ is large, one has the ``decoupling" limit, where $h$
becomes in many ways rather similar to the standard model Higgs.  If $A \to h$,
$m(A)$ would not be not too large and hadron colliders can search for the $A$
in the modes $A \to \tau^+\tau^-$, $bA \to b\tau^+\tau^-$ and $bA \to bbb$.
The $Abb$ and $A\tau\tau$ couplings are enhanced relative to the experimentally
difficult $Att$ and $A\nu\nu$ couplings by a factor of $\tan^2 \beta$ and so
limits on the maximum possible value of $\tan \beta$ can be set as a function
of $m(A)$.  John Conway and Flera Rizatdinova in this conference have
discussed the recent TeVatron results.  At this time, values of $\tan \beta$
over $\simeq 30$ are ruled out~\cite{CDF_MSSM,D0_MSSM} at $m(A) \simeq 130\gev$;
if these results are scaled by the expected final Run II luminosity and
$\tan^2 \beta$, it is reasonable to guess that the TeVatron experiments will
ultimately be able to set limits as low as $\tan \beta \simeq 20$.  More
detailed studies of the potential reach of the TeVatron and the LHC have been
done recently~\cite{MSSM_future}.

\section{NMSSM Higgs
		\label{sec:NMSSM_Higgs}}

Given the increasing restrictions on the available parameter space of the
minimal supersymmetric extension of the standard model, it is natural to
consider a nearly-minimal SUSY extension.  In the NMSSM SUSY model, the
smallest possible combination of fields is added to the known standard model
fields and their SUSY partners.  Neutral weak isospin singlet fermion and
corresponding complex scalar fields are introduced.  The resulting physical
content of the theory includes a new light pseudoscalar, $a$, which (in the
manner characteristic of Higgs bosons) decays into the heaviest kinematically
available particles.  For $m(a)$ above $2M_\mu$, $a \to \mu\mu$ is possible
and has a nearly 100\% branching ratio.  If $m(a)$ is over $\simeq 3$ times
the pion mass, hadronic decays become dominant; when $m(a) > 2 M_\tau$, the
decay into \tptm~becomes the dominant mode.  Interest in this model was
increased~\cite{HK_excitement} by the unusual dimuon mass spectrum observed in
$\Sigma \to p\mu^+\mu^-$ by the HyperCP~\cite{HK_thought} experiment.

In \epem~colliders, the $\Upsilon$ may decay to $a\gamma$ and there should be
a narrow peak in the $\gamma$ energy spectrum for events where a \tptm~or
\mpmm~pair has been identified.  A search using this method was performed
earlier by the CLEO collaboration~\cite{CLEO_no_a} which set limits on
\BR{$\Upsilon(1S)$}{$a\gamma$} ~$\times$ \BR{$a$}{$\mu^+\mu^-$} ~on the scale of
a few times $10^{-6}$ in the range of about $250\mev$ to $3.5\gev$, and also upon
\BR{$\Upsilon(1S)$}{$a\gamma$} ~$\times$ \BR{$a$}{$\tau^+\tau^-$} ~on the scale
of a few times $10^{-5}$ in the range of about $5$ to $9\gev$.  More
recently, BaBar~\cite{BaBar_no_a} examined their data for evidence of this
process, using the case where one $\tau$ decayed to $e \nu \overline\nu$ and
the other decayed to $\mu \nu \overline\nu$.  They set limits on
\BR{$\Upsilon(3S)$}{$a\gamma$} ~$\times$ \BR{$a$}{$\tau^+\tau^-$} ~on the scale
of a few times $10^{-5}$ in the range of about $4\gev$ to just under $10\gev$.

In a hadron collider, a pair of $a$ bosons would be produced as the result of
the decay of an $h$.  From LEP II, we have a very general
limit~\cite{OPAL_recoil} that the mass of any new scalar coupling to the $Z$,
including the $h$, must have a mass over $82\gev$, and so the $a$ is produced
in a hadron collider with a high boost.  That in turn means that its decay
into, say, a \mpmm~pair will produce particles with a small opening angle.  For
$m(a) < 2{\mathrm M}\tau$ the two tracks can be difficult to resolve in the
\rphi~plane.  D0~\cite{D0_NMSSM} has searched for the $a$ in the case
$2{\mathrm M}\tau < m(a)$ using the modes $aa \to \mu\mu\mu\mu$ and
$aa \to \mu\mu\tau\tau$.  The branching ratios are substantially lower than for
$aa \to \tau\tau\tau\tau$ but the signature is clearer.  Andy Haas has
discussed the special reconstruction criteria needed for these collinear
leptons in this conference.  Limits on 
$\sigma (p{\overline p} \to h)~\times$ \BR{$h$}{$aa$}
~of a few $\mathrm {pb}$ are obtained.

\section{Leptoquarks
		\label{sec:GUTschmutz}}

Because silicon vertex detectors can identify jets produced by fragmenting $b$
quarks, it is possibile to search for third generation leptoquarks at hadron
colliders.  An $LQ$-$\overline {LQ}$ pair would produce events containing 2
$b$ jets and a large \MET from the 2 $\nu_{\tau}$.  As Sergey Uzunan described
at this conference, this is the same signature as that which one might expect 
from pair production of ${\tilde b}$, with subsequent ${\tilde b} \to b \chi^0$
decays.  Limits can then be set~\cite{TwoForOne} upon both models as a result
of what is basically a single search method.  As a search for $\tilde b$,
limits up to $m(\tilde b) > 250\gev$ are obtained; as a search for leptoquarks,
$m(\mathrm {LQ}_3) > 252\gev$ is obtained.

The best way to find a leptoquark, at least a first generation one, is to take
a lepton and accelerate it to high energy and then arrange for it to collide
with a quark, similarly accelerated.  This is exactly what HERA did, collecting just
under $0.8\fb$ of $e^\pm p$ data at $\sqrt{s} = 300-319\gev$, $0.3\fb$ of which
had polarized $e^\pm$.  The ZEUS~\cite{ZEUS_LQ} collaboration measured the
$Q^2$ distribution in their data and compared it to the standard model
prediction.  The (very small) difference was then compared against deviations
that would be created by first generation leptoquarks, resulting in limits on
$m(LQ) / \lambda(LQ)$ of $0.5 - 1.9 \tev$, where $\lambda(LQ)$ is the coupling
of the leptoquark to the fermions.  Using the same technique they were also
able to set limits on large extra dimensions and contact interactions with the
same $Q^2$ distribution.  The H1 collaboration worked with different
kinematic variables, specifically, $M$ and $y$; their results~\cite{H1_LQ} are
not straight lines on the $\lambda(LQ)$ \vs~$m(LQ)$ plane.  If the couplings
are taken to be $\lambda(LQ) = \sqrt{4\pi\alpha_{em}}$, the H1 analysis rules
out leptoquark masses below 275 to $325\gev$, depending on the type of
leptoquark.

\section{Hidden Valley Scenarios
		\label{sec:DidYouFindItYet}}

As my long time friend and one of our kind hosts here in Detroit Dave Cinabro
once accurately pointed out, ``When somebody writes a paper that says he looked
for something and he did not find it, well then, you have to believe him."
Another, more common, reaction to a null search is to imagine that the imagined
new phenomena still actually does in fact exist, but at some higher energy
scale which is at least for the moment experimentally inaccessible.  Hidden
Valley scenarios are predicated on a third possible response: the new phenomena
still does exist at a relatively low mass scale, but is so weakly coupled to
the standard model phenomenology as to render it invisible, or at least, hard
to see.

One can postulate a wide range of fields that could exist in such a hidden
sector; ``hidden valleys" is really a class of models rather than than a
specific model.  In the simplest example of such a model~\cite{HV_theory} 
the valley is populated with two electrically neutral quarks which are
confined into so-called ``v-hadrons".  Some of these particles may be stable,
providing dark matter candidates; big-bang nucleosynthesis considerations
suggest that at least one v-hadron has to have a lifetime much less than 1 sec.
A $Z'$ that couples to both the hidden valley particles and the standard
model ones is included in this model, with a mass in the $1 \sim 6\tev$ range.

Andy Haas, in this conference, has presented D0's search~\cite{D0_HV} for
v-hadrons that are produced by mixing with a Higgs boson and have a long
lifetime; their decay is mediated by the $Z'$ and produces a pair of $b$ jets
that emanate from a vertex that is between 1.6 and $20\cm$ distant from the
\ppb~interaction point.  The large background from material interactions is
suppressed by comparing the locations of the jet vertices with the known
material distribution in the detector.  Limits on
$\sigma (p{\overline p} \to HX)~\times$ \BR{$H$}{$HV {\overline{HV}}$}
							   ~$\times~ \mbox{Br}^2(HV \to b{\overline b})$
as low as $1\,\mathrm {pb}$ are obtained.

\section{Supersymmetric Hidden Valley Dark Matter Model
		\label{sec:TheWholeBallOfWax}}

In recent years, a number of experiments have reported results that could be
interpreted as dark matter annihilation to \epem~pairs near the center of the
Milky Way.  Additionally, the DAMA experiment reports an annual modulation in
their NaI(Tl) detector which may be interpreted as a signal from a dark
matter galactic halo.  While there is no shortage of more mundane
explanations for these results, some authors~\cite{AH_etal} have taken a more
adventuresome approach.  They begin with the assumption that all of these
results are in fact due to new physics and then ask what would that new physics
look like.

\begin{figure}[h]
\centering
\includegraphics[width=80mm]{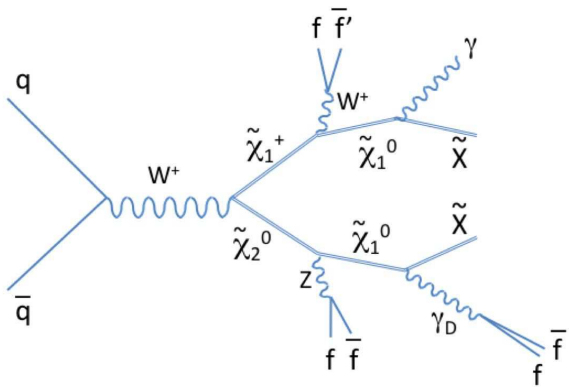}
\caption{A dark photon production diagram in \ppb~collisions.}
\label{Fig:ballOwax}
\end{figure}

They come to the surprising conclusion that dark matter is on the
$0.5 - 0.8 \tev$ mass scale and that it annihilates to standard model
particles with ``sizeable" cross-sections.  With such a large mass, it is
natural to speculate that a new symmetry prevents the rapid decay of such
states.  However, these states might couple to light (${\mathcal O} (1\gev)$)
particles, known as ``dark photons" $(\gamma_D)$.  They also have found that
such a picture can be implemented in a SUSY framework with GMSB.  In that case
a clear signature for \ppb~collider searches occurs through processes such as
that shown in Figure~\ref{Fig:ballOwax}; a high energy $\gamma$ would appear in
conjunction with \MET and a collinear \mpmm~pair from the $\gamma_D$ decay.

The low mass, high boost and decay into \mpmm~pairs of the dark photon means
that one may use the same reconstruction techniques as were applied in searching
for the NMSSM $a$ in hadron colliders.  The D0 collaboration has set
limits~\cite{D0_ballOwax} on $m({\tilde \chi}^0)$ as a function of $m(\gamma_D)$
in the range $0.1 < m(\gamma_D) < 2.5\gev$

\section{Model Independent Searches
		\label{sec:BruceGoneNow}}

Much of the motivation for searching for new physics beyond the standard model
stems from our dissatisfaction with the many aspects of the standard model which
we find so surprising.  Indeed, were it not for such astonishments as parity
violation, the $J/\psi$ observation, the large value of \Mt~and many others, the
standard model would surely have been easier to figure out!  While we do hope
and expect that getting the correct extension to the standard model will somehow
reduce our overall level of astonishment, history warns us that such an outcome
is not at all certain.  With this in mind, it behooves us to try to conduct
searches for new physics without the guidance of models that are at least in
part constructed so as to reduce our astonishment.

The basic scheme for the modern model-independent search begins by defining a
large number of final states.  The definition is usually made in terms of the
particle content of the final state, where particles are defined by the
detection capabilities of the experiment's apparatus.  So for example, final
states with low \PET~electrons would typically evade detection in hadron
colliders, and such final states can not be included.  Particles that require
unusual reconstruction schemes are typically not included in the list of
possible final states.  Particles that are found by vertexing their decay
products (such as \Ks~or \Dsp) have by and large not been included to date,
although there is no specific reason why they could not be.  One consequently
should not think of a model-independent search as being exactly the same as a
search for ``everything"; it is not quite that, at least to date.

For each entry on the list of possible final states, the standard model
processes contributing to the final state are identified and modeled.  The data
are then compared against this predicted background, and cases where the data
appear at a higher rate than the known physics rate are flagged.  Cases where
the data appear at a lower rate are also interesting, both as a check on the
method and in case there might be new physics amplitudes that interfere 
destructively with known amplitudes.  In assessing the statistical significance
of any departure of reality from prediction, it is important to allow for the
fact that the more comparisons you make, the more likely it is that the most
discrepant result will be at or beyond any particular level of significance.

There are different ways to compare the data to the predicted rates of
known physics.  There might be a different total number of events.
Distributions of kinematic variables for the data and the expectation can be
compared with an overall quality of fit statistic, such as the Komolgov-Smirnov
statistic.  The distribution of a kinematic variable, such as a reconstructed
mass, can be scanned for bumps.  Or one might scan the distributions of $\gev$
dimensioned kinematic variables such as \PET~or reconstructed mass from low
to high values, and look for discrepancies in the event counts above the scan
point.

This type of analysis has been completed at the CDF~\cite{CDF_mis},
D0~\cite{D0_mis} and H1~\cite{H1_mis} experiments, although not all three have
utilized the full range of possible comparison methods.  Jim Linnemann, in this
conference, has presented the D0 model independent search.
Table~\ref{Tab:MIScounts} shows the results of comparisons at the level of
simple event count comparisons of data with expected background levels.  The
H1 collaboration chose to express their results in terms of number of seen
events \vs~the expected backgrounds; to facilitate comparison with the CDF
and D0 results I have calculated a corresponding number of standard deviations.

\begin{table}[h]
\begin{center}
\caption{Significance of event count discrepancies in 3 model independent
searches.  See text regarding treatment of H1 results.\\}
\begin{tabular}{|c|c|c|} \hline
\textbf{CDF ($2.0\fb$)}                  &   \textbf{H1 ($0.5\fb$)}                & \textbf{D0 ($1.1\fb$)}                  \\
\hline
$\gamma \tau$ \hspace{5.5mm} $2.2\sigma$ &  $\nu 4j$ \hspace{4.0mm} $<3.0\sigma$   & $\mu jj$\MET \hspace{5.0mm} $9.3\sigma$ \\
\hline
$\mu \tau$    \hspace{5.5mm} $1.7\sigma$ &  $e   4j$ \hspace{4.0mm} $<2.4\sigma$   & $\mu j\gamma$\MET \hspace{5.0mm} $6.6\sigma$ \\
\hline
$e \tau$\MET  \hspace{2.5mm} $1.7\sigma$ &  $eee$ \hspace{4.0mm} $\sim2.0\sigma$   & \mpmm~\MET   \hspace{1.5mm} $4.4\sigma$ \\
\hline
                                         & $\mu\nu$ \hspace{4.0mm} $\sim1.5\sigma$ & \mpmm~$\gamma$ \hspace{3.0mm} $4.4\sigma$ \\
\hline
\end{tabular}
\label{Tab:MIScounts}
\end{center}
\end{table}

The statistically significant deviations in the channels flagged by the D0
analysis are attributed to defects in the modeling of the rate at which jets
fake as photons, trigger simulation shortcomings, and \PET~resolution effects
in the D0 tracking system which effect muon measurement.

Significantly, there is no overlap in the channels found by all 3 experiments.

%
%

\bigskip
\begin{acknowledgments}
I would like to thank a number of people who helped improve this presentation:
Todd Adams, Arnaud Duperrin, Andy Hass, Katjia Kruger, Monica D'Onofrio,
Monica Turcato, Stefan Schmitt and Tom Wright.  And I would also like very
much to thank our hard-working conference organizers for this very productive
meeting and for their gracious hospitality.
\end{acknowledgments}

\bigskip 

\end{document}